\documentstyle[12pt]{article}
\newcommand{\be}{\begin{equation}}
\newcommand{\ee}{\end{equation}}
\newcommand{\bea}{\begin{eqnarray}}
\newcommand{\eea}{\end{eqnarray}}
\newcommand{\ba}{\begin{array}}
\newcommand{\ea}{\end{array}}
\newcommand{\beas}{\begin{eqnarray*}}
\newcommand{\eeas}{\end{eqnarray*}}
\newcommand{\bes}{\begin{equation*}}
\newcommand{\ees}{\end{equation*}}

\begin{document}

\title{\bf Thermodynamics of Schr\"{o}dinger black holes with hyperscaling violation}
\author{J. Sadeghi \thanks{Email:pouriya@ipm.ir}\hspace{1mm}, B. Pourhassan \thanks{Email:b.pourhassan@umz.ac.ir}\hspace{1mm}, F. Pourasadollah \thanks{Email:f.pourasadollah@gmail.com}\\
{\small {\em  Sciences Faculty, Department of Physics, Mazandaran University,}}\\
{\small {\em P.O.Box 47416-95447, Babolsar, Iran}}}
 \maketitle

\begin{abstract}
In this work, we follow Kim and Yamada (JHEP1107 (2011) 120) and utilize AdS in light-cone frame to derive thermodynamic and transport properties of two
kinds of Schr\"{o}dinger black holes with hyperscaling violation. In that case, we show entropy and temperature are depend on $\theta$. In $\theta=0$ we
see our results are agree with the work of Kim and Yamada.
We also construct R-charged black hole with hyperscaling violation and obtain thermodynamics and transport properties.
\end{abstract}
\section{Introduction}
As we know the AdS/CFT correspondence is a powerful tool to study strongly coupled field theory [1]. Recently,  the extension of AdS/CFT in
non-relativistic with dynamical exponent $z\neq1$ [2-4] in the context of Schr\"{o}dinger  holography  for Lifshitz background studied by Ref. [5]. In the
recent years, various classes of background in application of condensed matter is discussed by the holographic method. In that case, the simple
generalization is to consider metric background  as  dual to scale-invariant field theories and non-conformal invariance [6]. So, we are going to introduce
the metric background as,
\begin{equation}
 ds ^{2} = -\frac{dt^{2}}{r^{2z}}+\frac{(dr^{2}+dx_{i}^{2})}{r^{2}},
\end{equation}
which is invariant under the following scaling,
\begin{equation}
t\rightarrow \lambda^{z}t  \qquad   , \quad x_{i}\rightarrow \lambda
x_{i} \quad ,\qquad r\rightarrow \lambda r.
\end{equation}
In order to describe gravity as a toy model of condensed matter systems, we need to generalize the AdS background. Recently, Lifshitz-type theories with
hyperscaling violation including an abelian gauge and scalar dilaton  field  discussed and the corresponding metric will be as,
\begin{equation}
ds ^{2} = r^{-2+ \frac{2 \theta}{d}}\left[-\beta r^{-2(z-1)}
dt^{2}+dr^{2}+ \sum_{i=1}^{d} dx_{i}^{2}\right].
\end{equation}
We note here the metric background includes a dynamical critical exponent $z$ and a hyperscaling violation exponent $\theta$, also $d$ is the number of
transverse dimensions. This metric is not scale invariant [6, 7], but transforms as,
\begin{equation}
ds\rightarrow \lambda ^{\frac{\theta}{d}}ds.
\end{equation}
Hyperscaling is a feature of the free energy based on naive dimension. For the theories with hyperscaling, the entropy behaves as $ S\sim T^{d/z}$ where
$T$, $d$ and $z$ are temperature, number of spatial dimensions and dynamical exponent respectively. Hyperscaling violation  first mentioned in context of
holographic in Ref. [8]. In this context, the hyperscaling violation exponent $ \theta$ is related to the transformation of the proper distance, and its
non-invariance implies the violation of hyperscaling of the dual field theory . Then, the relation between the entropy and temperature has been modified as
$ S\sim T^{(d-\theta)/z}$. In general, theory with hyperscaling violation $d-\theta$ plays the role of an effective space dimensionality for the dual field
theory. Inspired by the recent developments, we introduce Schr\"{o}dinger-type systems with Galilean invariance for arbitrary dynamical exponent $z$ with
hyperscaling violation exponent $\theta$ . The metric for $(d+3)$-dimensional Schr\"{o}dinger-type theory is given by,
\begin{equation}
ds ^{2} = r^{-2+\frac{2\theta}{d}}\left[-\beta r^{-2(z-1)} dt^{2}+dr^{2}
-2dtd\xi+ \sum_{i=1}^{d} dx_{i}^{2}\right].
\end{equation}
For maintaining the Galilean boost, existence an off-diagonal
component is necessary. The metric (5) is invariant under the
translations, rotations and Galilean boost, which is   following
form,
\begin{equation}
\vec{x}\rightarrow\vec{x}-\vec{\nu}t\quad ,\quad \acute{\xi}= \xi+
\frac{1}{2}(2\vec{\nu} .\vec{x}-\nu^{2}t),
\end{equation}
where the vectors are $d$-dimensional and have the following scale transformation,
\begin{eqnarray}
t&\rightarrow &\lambda^{z}t,\nonumber\\
\xi &\rightarrow &\lambda^{2-z}\xi, \nonumber\\
\vec{x}&\rightarrow &\lambda \vec{x}, \nonumber\\
r&\rightarrow &\lambda r.
\end{eqnarray}
The metric  background (5) is not invariant, but transforms covariantly $ ds\rightarrow \lambda ^{\theta / d}ds$ which is a defining property of
hyperscaling violation in holographic language. In general we can say that theories with hyperscaling violation has a dimensionful scale that dose not
decouples in the infrared. For this purpose, we use proper powers of this scale which will be denoted by $r_{F}$ to restore naive dimensional analysis.
Also note here, we will follow an effective holographic approach in which the dual theory lives on a finite $r$ slice.  Thus, the metric background (5)
provides a good description of the dual field theory only for a certain range of $r$, presumably for $ r\geq r_{F}$ anticipating the applications at the
low energy regions. The above information lead us to particularly work on Schr\"{o}dinger-type systems with hyperscaling violation by considering finite
temperature generalizations on $\beta=0 $ case. There exist a viable candidate for geometric realization of the Schr\"{o}dinger holography, called AdS in
light-cone frame ALCF [3, 4, 9] for the case $\beta=0$. The Schr\"{o}dinger symmetry can be geometrically realized in the pure $AdS$ space without a
deformation which are discussed by Goldberger [3],  Barbon and Fuertes [4]. They projected the theory onto a fixed momentum in one of the light-cone
directions and identified  the other light-cone coordinate as the time of the resulting non-relativistic theory. In addition to these two important
operations, the Refs. [9, 10] used light-cone coordinates with particular  normalization even for the zero temperature case as,
\begin{eqnarray}
x^{+}&=&b(t+x),\nonumber\\
x^{-}&=&\frac{1}{2b}(t-x).
\end{eqnarray}
So, we give first boost in the $x$-direction with rapidity $\log{b}$ and then go to the light-cone coordinates. The parameter $b$ exactly matches the extra
parameter generated in the Null-Melvin-Twist. Also we assign $b=-1$ so that the dynamical exponent $z=2$, so we have $[x^{+}]=-2$ and $[x^{-}]=0$.
Therefore,  in order to have the  non-relativistic we have to do scaling dimensions of light-cone coordinates, momentum projection and re-interpretation of
time. To express more clearly, a possible dual of a non-relativistic field theory can be pure AdS space in light-cone coordinates with the three mentioned
procedures. The hyperscaling violation already studied for various goals. For example, in the Ref. [11] the hyperscaling violation for scalar black holes
studied. Also, Ref. [12] studied thermodynamics of string black hole with hyperscaling violation, and Lifshitz metric with hyperscaling violation have been
studied in the Ref. [13]. In the interesting work [14], strongly coupled theories with hyperscaling violation have been studied by using holography, where
some important quantities such as Wilson loop and drag force [15] calculated. In this paper, we follow the direction of Ref. [10] and utilize ALCF
procedure for driving thermodynamic and transport properties of two kinds of Schr\"{o}dinger black hole solutions with hyperscaling violation. In section
2, we derive thermodynamics of planar Schr\"{o}dinger black hole with hyperscaling violation for the case of $ \beta=0$ and $z=1$. We show that all
thermodynamics properties of system change with $\theta$ and in limit of $\theta=0$ we see that our results are identical to Ref. [10]. In section 3, we
look for metric background that be solution of $R$-charged black hole with hyperscaling violation.  In order to have a physically reasonable theory, we
impose null energy condition so that in $ r=r_{F}$ we receive some condition on $\theta$. Next, we obtain thermodynamic and transport properties for that
black hole. Once again, we observe these properties change with $\theta$, and in the limit of $\theta=0$ they are identical to the results of Ref. [10].
\section {Planar black holes with hyperscaling violation}
In order to discuss thermodynamics of a black hole solution with hyperscaling violation, we use the method of Refs. [10, 16] for planar $AdS_{d+2}$ black
hole in the light-cone coordinates with the hyperscaling violation. In that case the corresponding ($d+2$)-dimensional action of gravitational system is
given by [17],
\begin{equation}
S=-\frac{1}{16\pi G} \int d^{d+2}x\sqrt{-g}\left[R-\frac{1}{2}(\partial\phi)^{2}+V(\phi)\right]-\frac{1}{8\pi G}\int d^{d+1}x
\sqrt{-\gamma}(K+\frac{3}{R}+\frac{R}{4}\Re),
\end{equation}
where $\phi$ is dilaton field and the symbols $g$, $\Re$ and $R$ are determinant of the metric,  scalar curvature and  length scale of  the theory,
respectively. Boundary terms are also included in (9) and $\gamma $ denotes the determinant of boundary metric.  The first term is Gibbons-Hawking term
[18] with the trace of the boundary second-fundamental form $K$ and the second and third terms are standard five-dimensional counterterms of
Balasubramanian and Karus [19] with the (intrinsic) scalar curvature of boundary $\Re$. The black hole solution with hyperscaling violation from the action
(9) for $(d+2)$-dimensional case can be written as [6, 17],
\begin{eqnarray}
d s^{2}_{d+2}&=& (\frac{r_{F}}{r})^{\frac{2\theta}{d}}\{(\frac{r}{R})^{2}[-hdt^{2}+d x_{i}^{2}]+({\frac{R}{r}})^{2}h^{-1}dr^{2}\},\nonumber\\
h&=&1-\frac{r_{H}^{d+1-\theta}}{r^{d+1-\theta}},
\end{eqnarray}
where  $i=1,...,d$,  $r=r_{H}$ is  location of horizon and $r=r_{F}$ is boundary. We should here note that the line element (10) is exactly metric of the
Ref. [6] under transformation $r\sim1/r$ and setting $z=1$. As mentioned before, we use the light-cone coordinates as,
\begin{equation}
x^{+}=b(t+x_{1}) \quad, \quad x^{-}=\frac{1}{2b}(t-x_{1}).
\end{equation}
In this case the metric  background will be as,
\begin{equation}
d s^{2}_{d+2}= (\frac{r_{F}}{r})^{\frac{2\theta}{d}}\left[(\frac{r}{R})^{2}\left(\frac{1-h}{4b^{2}}d x^{+2}+ (1-h)b^{2}dx^{-2}-(1+h)dx^{+}dx^{-}+d
x_{j}^{2}\right)+(\frac{R}{r})^{2}h^{-1}dr^{2}\right],
\end{equation}
where $j=2,..., d$.  In order to have the  non-relativistic limit we re-interpret one of the light-cone coordinates $x^{+}$ as the time. So, the ADM form
of metric background is,
\begin{equation}
d s^{2}_{d+2}= (\frac{r_{F}}{r})^{\frac{2\theta}{d}}\left[(\frac{r}{R})^{2}\left(\frac{-h}{1-h}b^{-2}dx^{+2}+
(1-h)b^{2}(dx^{-}-\frac{1}{2b^{2}}\frac{1+h}{1-h}dx^{+})^{2}+d x_{j}^{2}\right)+(\frac{R}{r})^{2}h^{-1}dr^{2}\right].
\end{equation}
By using the above metric, one can obtain  $N$ (laps function), $V^{i}$ (shift function) and the horizon velocity (which can be interpreted as chemical
potential) in the unit of $x^{-}$ as,
\begin{eqnarray}
N&=&\frac{r_{F}^{\frac{\theta}{d}} r^{1-\frac{\theta}{d}}}{b}\sqrt{\frac{h}{1-h}},\nonumber\\
V^{-}&=&-\frac{1}{2b^{2}}\frac{1+h}{1-h},\nonumber\\
\Omega_{H}&=&\frac{1}{2b^{2}}.
\end{eqnarray}
For calculating the projections, we obviously need  to construct the
following normal vectors to the surfaces,
\begin{eqnarray}
n_{\mu}={\sqrt{g_{rr}}}=\frac{R}{r}h^{-\frac{1}{2}}(\frac{r_{F}}{r})^{\frac{\theta}{d}}(0,0,0,0,1)
,\quad u_{\mu}=-N(1,0,0,0,0)
\end{eqnarray}
where $n_{\mu}$ and $u_{\mu}$ are timelike  and spacelike boundary
surface normals with the components ordering (+, -, 2, 3, ...,$d$,
$r$) respectively. So, the projections in four dimensional timelike
boundary and three-dimensional intersection surface are given by,
\begin{equation}
\gamma_{\mu\nu}=g_{\mu\nu}-n_{\mu}n_{\nu} \qquad , \qquad
\sigma_{\mu\nu}=g_{\mu\nu}-n_{\mu}n_{\nu}+u_{\mu}u_{\nu}.
\end{equation}
At the first, we use stress-energy-momentum tensor and obtain the thermodynamic properties of system. In that case the corresponding stress-energy-momentum
tensor will be as,
\begin{equation}
T_{ij}=\frac{-2}{\sqrt{-\gamma}}\frac{\delta I_{bdy}}{\delta \gamma^{ij}}\frac{1}{8\pi
G_{d+2}}(K_{ij}-\gamma_{ij}K-\frac{d}{R}\gamma_{ij}+\frac{R}{2}G_{(d+1)ij}),
\end{equation}
where $K_{\mu\nu}=-n_{\nu;\mu}$, and $G_{(d+1)ij}$ is the Einstein tensor with respect to the metric $ \gamma_{ij}$ and also $ i,j\in \{+,-,2,...,d\}$. The
$T_{ij}$ evaluated at the boundary of the solution (12) and then the boundary is moved to $ r_{F}.$ The energy-momentum tensor (17) is used to obtain total
mass,
\begin{equation}
M=\int d^{d}x\sqrt{\sigma}(N\epsilon-V^{a}j_{a}),
\end{equation}
where
\begin{eqnarray}
\epsilon&=&u_{i}u_{j}T^{ij},\nonumber\\
j_{a}&=&-\sigma_{ai}u_{j}T^{ij},
\end{eqnarray}
with $a,b\in \{-,2,...,d\}$. Also one can obtain total particle number,
\begin{equation}
J=\int d^{d}x\sqrt{\sigma}j_{a}\phi^{a},
\end{equation}
where $\phi^{a}=(1,0,0,...0)$ is the killing vector filed. The entropy of system is given by,
\begin{equation}
S=\frac{A}{4\pi G_{d+2}}.
\end{equation}
As we know the temperature can be derived through the computation of surface gravity [20]. According to equation (17), for metric background (12), we have
following expressions of the surface gravity,
\begin{equation}
K_{++}=\frac{-h^{\frac{1}{2}}(\frac{-2\theta}{d}+1+\theta-d)r_{F}^{\frac{\theta}{d}}r_{H}^{d+1-\theta}r^{\frac{-\theta}{d}+1+\theta-d}}{8R^{d}b^{2}},
\end{equation}
and
\begin{equation}
K_{+-}=\frac{h^{\frac{1}{2}}r_{F}^{\frac{\theta}{d}}}{R^{d}}\left[(1-\frac{\theta}{d})r^{2-\frac{\theta}{d}}-\frac{(\frac{-2\theta}{d}+1+\theta-d)}{4}
r_{H}^{d+1-\theta}r^{-\frac{\theta}{d}+1+\theta-d}\right].
\end{equation}
So, one can write $M$ and $J$ in terms of $ K_{+-}$ and $K_{++}$  as,
\begin{equation}
J=\frac{-V_{d}r_{F}^{\theta}r^{-\frac{\theta}{d}+1+\theta-d}b^{2}}{8\pi
G_{d+2} R^{d-1}h^{\frac{1}{2}} r_{F}^{\frac{\theta}{d}}}[2(1+h)K_{++}-(1-h)K_{+-}],
\end{equation}
and,
\begin{equation}
M=\frac{-V_{d}u_{+}r_{F}^{\frac{2\theta}{d}}r^{2(1-\frac{\theta}{d})}}{8\pi G_{d+2} R^{d+2}}
\left[-2(1-h)K^{++}+\frac{(1+h)}{2}K^{+-}-\frac{h}{1-h}\gamma^{++}(K+\frac{d}{R})\right],
\end{equation}
where,
\begin{equation}
\gamma^{++}=\frac{-(1-h)R^{2}}{hr_{F}^{\frac{2\theta}{d}}r^{2(1-\frac{\theta}{d})}},
\end{equation}
and,
\begin{equation}
K=\frac{-(\frac{r_{F}}{r})^{\frac{\theta}{d}}h^{\frac{1}{2}}}{R}\left[(\frac{-2\theta}{d}+1+\theta-d)\frac{1-h}{2h}-(1-\frac{\theta}{d})(\frac{1}{h}+d)\right].
\end{equation}
Eliminating components of surface gravity gives the thermodynamical quantities of planar Schr\"{o}dinger black hole with hyperscaling violation as the
following,
\begin{equation}
S=\frac{V_{d}r_{F}^{\theta}r_{H}^{d+1-\theta}b}{4 G_{d+2} R^{d}}
\:\:\:\:,\:\:\:\:\beta=\frac{4\pi R^{2}b}{r_{H}(d+1-\theta)},
\end{equation}
and,
\begin{equation}
J=\frac{-V_{d}r_{F}^{\theta}r_{H}^{d+1-\theta}b^{2}}{4\pi
G_{d+2} R^{d+2}}[\frac{d+1-\theta}{4}],
\end{equation}
Also,
\begin{equation}
M=\frac{V_{d}}{8\pi G_{d+2} R^{d+2}}\left[\frac{(d-3\theta-1)}{4}r_{F}^{\theta}r_{H}^{d+1-\theta}-\theta r_{F}^{d+1}\right].
\end{equation}
It is important to note that in case of  $\theta=0$ and $d=3$ we receive the results of Ref. [9] for planar Schr\"{o}dinger black hole without hyperscaling
violation. Therefore, we success to obtain thermodynamical quantities in presence of hyperscaling violation. We find that the new entropy has an additional
factor $(\frac{r_{F}}{r_{H}})^{\theta}$. So, because of $r_{F}>r_{H}$, one can find that the effect of hyperscaling violation is increasing of the entropy.
Now we are going to investigate thermodynamical properties of system with the variation of $\theta.$ First we consider $\theta=d-1 $ , so we have,
\begin{equation}
S=\frac{V_{d}r_{F}^{d-1}r_{H}^{2}b}{4 G_{d+2} R^{d}}\:\:\:\:,\:\:\:\:\beta=\frac{2\pi R^{2}b}{r_{H}},
\end{equation}
and
\begin{eqnarray}
J&=&\frac{-V_{d}r_{F}^{d-1}r_{H}^{2}b^{2}}{8\pi G_{d+2} R^{d+2}}\nonumber\\
M&=&\frac{V_{d} (1-d)}{8\pi G_{d+2} R^{d+2}}[r_{F}^{d-1}r_{H}^{2}+r_{F}^{d+1}].
\end{eqnarray}
At the second, we consider $\theta=d$ and obtain,
\begin{equation}
S=\frac{V_{d}r_{F}^{d}r_{H}b}{4 G_{d+2} R^{d}}\:\:\:\:,\:\:\:\:\beta=\frac{4\pi R^{2}b}{r_{H}},
\end{equation}
and
\begin{eqnarray}
J&=&\frac{-V_{d}r_{F}^{d}r_{H}b^{2}}{16\pi G_{d+2} R^{d+2}}\nonumber\\
M&=&\frac{-V_{d}} {8\pi G_{d+2} R^{d+2}}[\frac{(2d+1)r_{F}^{d}r_{H}}{4}+dr_{F}^{d+1}].
\end{eqnarray}
For the $\theta=d+1$ the corresponding $h$ will be zero, so in this case we don't have any particles. It is good to remind that the range of $d-1\leq
\theta\leq d$ is of our interest [6-7],  so it is reasonable for the $\theta=d+1 $ case to be invalid. In $\theta=(d-1)/3 $, mass is independent of the
location of horizon and it gives the negative value. It may be interesting to involve the cases with negative $\theta$, in that case there are some
difficulties to analyze energy scale. In the next section we will  derive the deformed R-charged black hole metric with hyperscaling violation, then we
impose null energy condition and compute the thermodynamical quantities in the $\theta\neq 0$ case.
\section{R-charged black holes with hyperscaling
violation} In this section, first we construct R-charged black hole background with hyperscaling violation, then discuss thermodynamics of system and
obtain entropy.
\subsection{R-charged black holes}
Generally, the R-charged black hole have three independent charges and charged black hole are static solutions of $N=2$ supergravity. The thermodynamic
properties for general charge configuration obtained by ALCF procedure in Ref. [9]. In this section initially we want to find the form of R-charged black
hole with hyperscaling violation, in that case we consider the bosonic part of the effective gauged supersymmetric $N=2$ Lagrangian which describes the
coupling of vector multiples to supergravity [20],
\begin{equation}
e^{-1}L=\left(\frac{\Re}{2}-\frac{1}{2}g_{xy}\partial_{\mu}\phi^{x}\partial^{\mu}\phi^{y}-\frac{1}{4}a_{IJ} F_{\mu\nu}^{I}F^{\mu\nu
J}-g^{2}V+\frac{e^{-1}}{48}\epsilon^{\mu\nu\rho\sigma\lambda}C_{IJK} F_{\mu\nu}^{I}F_{\rho\sigma}^{J}A_{\lambda}^{K}\right),
\end{equation}
where $e=\sqrt{-g}$ is the determinant of veilbein, $\Re\:$ is the Ricci scalar, $g_{xy}  $ is a metric on the scalar manifold, $a_{IJ}$ is a kinetic gauge
coupling of the field strength and $g$ is a constant gauge coupling, $g_{xy}, a_{IJ}$ and $V$ are functions of scalar field $\phi^{x}  $. The variation of
Lagrangian (35) with respect to $g_{xy}$ ,$\phi^{x}$  and  $F_{\mu\nu}^{I}$ gives the field equations of motion. So,  a solution of the equation help us to
take the following ansatz for the corresponding metric background,
\begin{eqnarray}
d s^{2}_{d+2}&=&-a(r)^{2}dr^{2}+b(r)^{2}d t^{2}+c(r)^{2}d s^{2}_{d,k}\nonumber\\
d s^{2}_{d,k}&=&\bar{g}_{ij}dx^{i}dx^{j}\:\:(i,j=1,2,...,d),
\end{eqnarray}
and the Ricci  tensor for the metric $\bar{g}_{ij}$   will be
following form,
\begin{equation}
\bar{R}_{ij}=k(d-1)\bar{g}_{ij}.
\end{equation}
The curvature constant $k$ can be positive, zero or negative corresponding to three different geometries of hypersurface: $k=1$ is the $d$-dimensional
sphere, $k=0$ is the d-dimensional flat space and $k=-1$ is the d-dimensional hyperbolic space. They can be described by following equation,
\begin{equation}
d s^{2}_{d}= \left\{
\begin{array}{ll}
d\theta^{2}+\sin^{2}\theta d\Omega_{d-1}^{2},  &  k=+1\\
d\theta^{2}+\theta^{2} d \Omega_{d-1}^{2},   &   k=0\\
 d\theta^{2}+\sinh^{2}\theta d \Omega_{d-1}^{2}, & k=-1\\
\end{array}
\right.
\end{equation}
We can solve equations of motion with the following relation,
\begin{equation}
\ln a+\ln b+\ln c+(3\frac{\theta}{d}-1)\ln r=(d-3)U,
\end{equation}
where
\begin{eqnarray}
\ln a&=&U-\frac{1}{2}\ln f-\frac{\theta}{d}\ln r,\nonumber\\
\ln b&=& -(d-1)U+\frac{1}{2}\ln f-\frac{\theta}{d}\ln r, \nonumber\\
\ln c&=&U+(1-\frac{\theta}{d})\ln r,
\end{eqnarray}
and also,
\begin{equation}
f=-k+\frac{\mu}{r^{d-\theta-1}}-g^{2}r^{2}e^{2dU}.
\end{equation}
By inserting these relations into equations of motion we have the following equation,
\begin{equation}
 U''+(d-1)U'^{2}+[d-\frac{\theta}{d}(d-2)]\frac{U'}{r}-\frac{\frac{\theta}{d}(\frac{\theta}{d}-1)}{r^{2}}=0.
\end{equation}
consequently for d+2=5, we have,
\begin{equation}
e^{2U}=H_{I}=(\frac{r_{F}}{r})^{\frac{\theta}{3}}+\frac{q_{I}}{r_{F}^{\frac{2\theta}{3}}r^{2(1-\frac{\theta}{3})}},
\end{equation}
and the form of  metric  background with hyperscaling violation can
be written as,
\begin{equation}
ds_{5}^{2}=(\frac{r_{F}}{r})^{2\frac{\theta}{3}}[fe^{-4U}dt^{2}-f^{-1}e^{2U}dr^{2}+r^{2}e^{2U}ds_{3}^{2}].
\end{equation}
We define following expressions,
\begin{equation}
e^{6U}=H=H_{1}H_{2}H_{3}=(\frac{r_{F}}{r})^{\theta}+\frac{Q_{1}}{r^{2}}+\frac{Q_{2}}{r_{F}^{\theta}r^{4-\theta}}+\frac{Q_{3}}{r_{F}^{2\theta}r^{6-2\theta}},
\end{equation}
where $Q_{1}=q_{1}+q_{2}+q_{3}$, $Q_{2}=q_{1}q_{2}+q_{2}q_{3}+q_{1}q_{3}$, and $Q_{3}=q_{1}q_{2}q_{3}$. So, similar to Ref. [10] one can rewrite the metric
(44) for R-charged black hole as,
\begin{equation}
ds^{2}=(\frac{r_{F}}{r})^{2\frac{\theta}{3}}\left[\frac{r^{2}}{R^{2}}H^{\frac{1}{3}}(-hdt^{2}+
\eta_{k}^{2}+d\chi_{k}^{2})+\frac{R^{2}}{r^{2}}H^{-\frac{2}{3}}h^{-1}\right],
\end{equation}
where
\begin{eqnarray}
h&=&1+(\frac{R}{r})^{2}\{k-(\frac{r_{0}}{r})^{2-\theta}\}H^{-1},\nonumber\\
f&=&\frac{r^{2}}{R^{2}}Hh,\nonumber\\
\mu &=&r_{0}^{2-\theta}.
\end{eqnarray}
We introduce  parameter $r_{H}$ as the location of horizon in the $r$ coordinate such that it is the largest root of $h=0$. According to the value of $k$
which introduced before, we have,
\begin{eqnarray}
\eta_{+1}&=&\frac{R}{2}(d\psi+\cos\theta d\phi),\nonumber\\
\eta_{-1}&=&\frac{R}{2}(d\psi+\cosh\theta d\phi),\nonumber\\
\eta_{0}&=&dx,
\end{eqnarray}
and $d\chi_{k}^{2}$ are identical to the relation (38). By using  the above information, we are ready to  derive thermodynamical properties for this kind
of black hole with hyperscaling violation with ALCF procedure. However, first we are going to investigate null energy condition and study some constraints.
After generating a particular solution with hyperscaling violation for R-charged black hole the physically sensible dual field theory give us motivation to
look for some constraints. In order to obtain such constraints on (46) in the presence of negative cosmological constant [21], the null energy condition
(NEC) must be satisfied [6,7].
\begin{equation}
T_{\mu\nu}N^{\mu}N^{\nu}\geq 0.
\end{equation}
The null vectors are revelent together with $N^{\mu}N_{\mu}=0$ and they have the following definition,
\begin{equation}
N^{t}=\frac{1}{\sqrt{h}H^{\frac{1}{6}}r^{1-\frac{\theta}{3}}},\quad N^{r}=\frac{\sqrt{h}\cos \phi}{H^{\frac{-1}{3}}r^{-1-\frac{\theta}{3}}},\quad
N^{i}=\frac{\sin\phi}{H^{\frac{1}{6}}r^{1-\frac{\theta}{3}}}\quad,
\end{equation}
where $\phi=0$ or $\pi/2 $. From this, we can get two independent null energy condition in the limit of $r=r_{F}$ as,
\begin{eqnarray}
\theta &\leq &0.365 \nonumber\\
\theta &\geq &1.315,
\end{eqnarray}
with $(1-\frac{\theta}{2})^{2}\geq 0$. The second condition is always maintained and the first gives new ranges rather than what derived before. However,
they are also realized in the sting theory construction.

\subsection{Thermodynamical quantities}
The thermodynamical quantities of the R-charged black holes with hyperscaling violation can be derived similar way of the section 2. The normal vectors of
hypersurfaces and the projection tensors are defined similarly as before. In order to obtain a metric in light-cone coordinate we proceed as Ref. [10] and
take following coframe,
\begin{equation}
\omega^{+}=b(dt+\eta_{k}),\quad\omega^{-}=\frac{1}{2b}(dt-\eta_{k}), \quad \omega_{k}=\omega^{r}=dr,
\end{equation}
where,
\begin{equation}
\vec{\omega}_{0}=(dy,dz),\quad\vec{\omega}_{+1}=\frac{R}{2}(d\theta,\sin\theta d\phi),\quad \vec{\omega}_{-1}=\frac{R}{2}(d\theta,\sinh\theta d\phi),
\end{equation}
are given by the Ref. [18]. Therefore our desired form of metric for the $\theta\neq0$ case is gained,
\begin{equation}
d s^{2}=(\frac{r_{F}}{r})^{\frac{2\theta}{3}}\left[(\frac{r}{R})^{2}H^{\frac{1}{3}}\left(\frac{1-h}{4b^{2}}\omega^{+2}+
(1-h)b^{2}\omega^{-2}-(1+h)\omega^{+}\omega^{-}+\vec{\omega}_{k}^{2}\right)+(\frac{R}{r})^{2}H^{-\frac{2}{3}}h^{-1}\omega^{r2}\right]
\end{equation}
So, the ADM form of the metric obtained as the following,
\begin{equation}
d s^{2}= (\frac{r_{F}}{r})^{\frac{2\theta}{3}}\left[(\frac{r}{R})^{2}H^{\frac{1}{3}}\left(\frac{-h}{1-h}b^{-2}\omega^{+2}+
(1-h)b^{2}(\omega^{-}-\frac{1}{2b^{2}}\frac{1+h}{1-h}\omega^{+})^{2}+\vec{\omega}_{k}^{2}\right)+(\frac{R}{r})^{2}H^{-\frac{2}{3}}
h^{-1}\omega^{r2}\right],
\end{equation}
By using above metric as before one can obtain,
\begin{eqnarray}
N&=&r_{F}^{\frac{\theta}{3}} \frac{r^{1-\frac{\theta}{3}}}{R}H^{\frac{1}{6}}\sqrt{\frac{h}{1-h}}b^{-1},\nonumber\\
V^{-}&=&-\frac{1}{2b^{2}}\frac{1+h}{1-h},\nonumber\\
\Omega_{H}&=&\frac{1}{2b^{2}}.
\end{eqnarray}
It is necessary to remark that our boundary action for this case is completely similar to the action (9), except that we must add the Liu-Sabra counterterm
[21] which is corresponding to the variation of the renormalization scheme in the dual field theory. So, we have,
 \begin{equation}
S_{boundary}=-\frac{1}{8\pi G_{5}}\int d^{4}x \sqrt{-\gamma}(K+\frac{3}{R}+\frac{R}{4}\Re_{4}+\frac{{\phi}^{2}}{2R}),
\end{equation}
where,
\begin{equation}
\phi^{2}=\frac{1}{r^{4}}(\frac{2}{3}(Q_{1})^{2}-2Q_{2}),
\end{equation}
is a new term which added to the boundary action of (9). We calculate stress-energy tensors from the on-shell value variation of the above boundary action.
For the cases $k=\pm1$ they should be computed in the non-coordinate bases [9] and the masses and momenta are obtained by the Brown-York procedure. The
entropy and temperature are obtained from the metric expressed in the above coframes. In addition, we are interested in deriving chemical potentials and
$N_{i}$ which is produced through the Gauss law,
\begin{equation}
N_{i}=\lim _{r\rightarrow r_{F}} \frac{-1}{16\pi
G_{5}}\int\omega^{\wedge
3}\sqrt{\sigma}n_{\mu}u_{\nu}F_{i}^{\mu\nu},
\end{equation}
and the chemical potentials are given by the first term of following equation [10],
\begin{equation}
A_{i}(r=r_{F})=\frac{g_{i}}{r_{H}^{2-\frac{\theta}{3}}r_{F}^{\frac{\theta}{3}}+q_{i}r_{H}^{2\frac{\theta}{3}}
r_{F}^{-2\frac{\theta}{3}}}\left[\frac{\omega^{+}}{b^{-1}}+b(\omega^{-}-\omega^{+}\frac{1}{2b^{2}})\right],
\end{equation}
where $g_{i}$ are $U(1)^{3}$ charges which are dependent on parameter $\theta$,
\begin{equation}
g_{i}=r_{F}^{\frac{\theta}{3}}\left(\frac{k\theta}{6}r_{F}^{4}+\theta
r_{F}^{2}-r_{0}^{2-\theta}\frac{\theta}{6}(r_{F}^{\theta-2}+\frac{r_{F}^{2}}{2})+q_{i}[kq_{i}(1-
\frac{\theta}{6})+r_{0}^{2}(1-\frac{7\theta}{6})]\right)^{\frac{1}{2}}.
\end{equation}
Hence we obtain the thermodynamical quantities as,
\begin{equation}
J=\frac{-V_{3}}{8\pi G_{5}R^{3}}\left[(2-\theta)r_{0}^{2-\theta}r_{F}^{\theta}+(1-\frac{\theta}{2})k(\frac{4}{3} Q_{1}-r_{F}^{2})\right],
\end{equation}
and,
\begin{equation}
M=\frac{V_{3}}{8\pi G_{5}R^{3}}\left[(\frac{1}{2}-\theta)r_{0}^{2-\theta}r_{F}^{\theta}+\frac{ (1-\theta)}{2} \frac{k}{3}Q_{1}-(1-\frac{5}{4}\theta)k
r_{F}^{2}+\frac{3\theta r_{F}^{4}}{2R^{2}}+\frac{4r_{F}^{2}}{3R^{2}}\theta Q_{1}+\frac{\theta Q_{2}}{2R^{2}}-\frac{3|k|R^{2}}{8}\right],
\end{equation}
also,
\begin{equation}
\mu_{i}=\frac{g_{i}}{r_{H}^{2-\frac{\theta}{3}}r_{F}^{\frac{\theta}{3}}+q_{i}r_{H}^{2\frac{\theta}{3}}r_{F}^ {-2\frac{\theta}{3}}}b^{-1},
\end{equation}
and,
\begin{equation}
N_{i}(r)=\frac{g_{i}V_{3}b(1-\frac{\theta}{3})}{8\pi G_{5}R^{3}}.
\end{equation}
Then, entropy is obtained as the following,
\begin{equation}
S=\frac{r_{H}^{3-\theta}H_{H}^{\frac{1}{2}}V_{3}b}{4G_{5}R^{3}},
\end{equation}
and inverse of temperature is given by,
\begin{equation}
\beta=\frac{\pi \sqrt{H(r_{H})}R^{2}b}{r_{H}Q_{k}},
\end{equation}
where
\begin{equation}
Q_{k}=(1-\frac{\theta}{2})\left[(\frac{r_{F}}{r_{H}})^{\theta}+\frac{Q_{1}+kR^{2}}{2r_{H}^{2}}-\frac{Q_{3}} {r_{H}^{6-2\theta}r^{2\theta}}\right].
\end{equation}
We can see that effect of hyperscaling violation on the entropy is a factor $r_{H}^{-\theta}$, therefore in the case of $r_{H}>1$ the hyperscaling
violation effect is decreasing entropy while in the case of $r_{H}<1$ the hyperscaling violation effect is increasing the entropy.\\
For the case of $r_{F}=R/\sqrt{2}$ the equations (62) and (63) give the following results,
\begin{equation}
J=\frac{-V_{3}}{4\pi G_{5}R^{3}}(1-\frac{\theta}{2})\left[r_{0}^{2-\theta}(\frac{R}{\sqrt{2}})^{\theta}+k\frac{2}{3} Q_{1}+\frac{|k|R^{2}}{4}\right],
\end{equation}
and
\begin{equation}
M=\frac{V_{3}}{16\pi G_{5}R^{3}}\left[(1-2\theta)r_{0}^{2-\theta}(\frac{R}{\sqrt{2}})^{\theta}+\frac{(1-5\theta)}{2} \frac{2k}{3}Q_{1}+\frac{\theta
Q_{2}}{2R^{2}}+\frac{|k|R^{2}(1-2\theta)}{4}\right].
\end{equation}
For $\theta=0$ we see that the results completely agree with Ref. [10] for R-charged black hole.
\section{Conclusion}
In this paper, we used Ref. [10] and obtained thermodynamical and transport properties of two kinds of Schr\"{o}dinger black holes with hyperscaling
violation. We have shown that all thermodynamical properties of system changed with $\theta$. In the first case we found that the hyperscaling violation
increased the entropy by $(\frac{r_{F}}{r_{H}})^{\theta}$. We confirmed that our results in $\theta=0$ limit are agree with the work of Kim and Yamada
[10]. We Also constructed the solution of R-charge black hole with hyperscaling violation and also imposed the null energy condition in $r=r_F$, so we
received some condition on the $\theta$. For the solution of R-charged black hole, also we obtained thermodynamical and transport properties depend on
$\theta$. In that case we found that modification of entropy due to hyperscaling violation is depend on $r_{H}^{-\theta}$. In the case of $\theta=0$ the
thermodynamical properties of
 system completely agree with Ref. [10].

\end{document}